\documentclass[aps,twocolumn,reprint, superscriptaddress,
  groupedaddress]{revtex4}

\usepackage{amsmath}    
\usepackage{graphicx}   
\usepackage{verbatim}   
\usepackage{color}      
\usepackage{subfigure}  
\usepackage{hyperref}   
\usepackage{amssymb}
\usepackage{epstopdf}
\usepackage{float}
\usepackage{xcolor}
\usepackage[normalem]{ulem}

\definecolor{red}{rgb}{0.6,0,0}

\begin{document}

\title{RNA Base Pairing Determines the Conformations of RNA Inside Spherical Viruses}
\author{Gonca Erdemci-Tandogan}
\email[email: ]{gonca.erdemci@email.ucr.edu}
\email[\\Present address: ]{Department of Physics, Syracuse University, Syracuse, NY 13244, USA}
 \affiliation{Department of Physics and Astronomy,
   University of California, Riverside, California 92521, USA}
 \author{Henri Orland}
 \affiliation{Institut de Physique Th$\acute{e}$orique, CEA-Saclay, CEA, F-91191 Gif-sur-Yvette, France}
 \affiliation{Beijing Computational Science Research Center, No.10 East Xibeiwang Road, Haidan District, Beijing 100193, China }
\author{Roya Zandi}
 \affiliation{Department of Physics and Astronomy,
   University of California, Riverside, California 92521, USA}

\begin{abstract}
Many simple RNA viruses enclose their genetic material by a protein shell called the capsid. While the capsid structures are well characterized for most viruses, the structure of RNA inside the shells and the factors contributing to it remain poorly understood. We study the impact of base pairing on the conformations of RNA and find that it undergoes a swollen coil to globule continuous transition as a function of the strength of the pairing interaction. We also observe a first order transition and kink profile as a function of RNA length.  All these transitions could explain the different RNA profiles observed inside viral shells.
\end{abstract}
\maketitle

The simplest viruses are built from a protein shell called the capsid that surrounds its genome (RNA or DNA) \cite{Bancroft}. Due to the electrostatic interactions, under many {\it in vitro} conditions, capsid protein (CP) subunits of many single-stranded RNA viruses assemble spontaneously around the genome or other negatively charged polymers, to form symmetric shells with extraordinarily monodisperse size distributions \cite{Zlotnick,Garmann2014,Comas2014,Lin2012}. These features have made viruses ideal for several material science and bionanotechnology applications such as gene therapy and drug delivery\cite{ZhangReview2015}.

\begin{figure}[ht]
  \includegraphics[width=7cm]{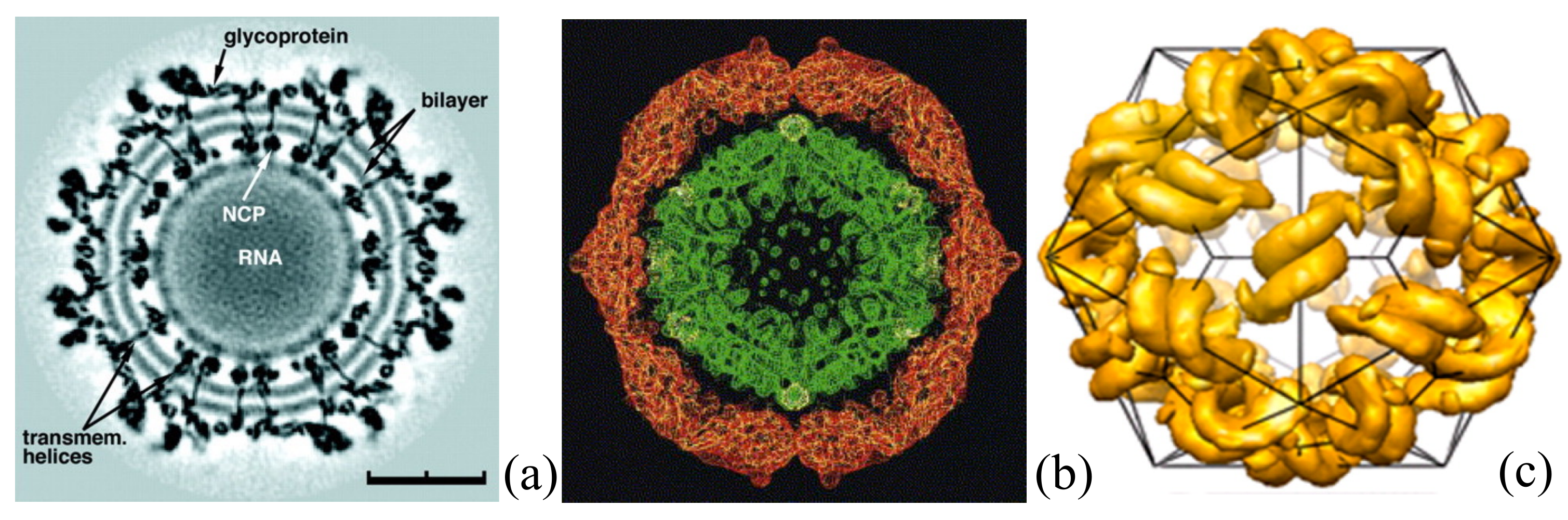}
 \caption{Image reconstruction of the Sindbis (a), FHV virus (b) and  STMV (c) obtained from cryoelectron micrographs from Refs. \cite{Zhang2002Sindbis, Tsuruta1998FHV, Schroeder2014STMV} respectively. Their relative sizes are not at scale.}
 \label{figure1} 
\end{figure}

While the structure of capsids for most viruses is completely understood from the cryoelectron microscopy (cryo-EM) or x-ray analysis \cite{BOTTCHER1997-HepB-CA,Yu2008-CPV-CA,Liu2016-Rhino-CA,Zhang2008-Rota-CA}, despite ongoing intense experimental studies \cite{Koning2016,Kuhn2002Dengue,Zhang2002Sindbis,Tsuruta1998FHV,Toropova2008MS2,Fox1998CCMV,Schroeder2014STMV,Hu2008,Zeng2012} the structure and the map density of RNA inside of viral shells and the factors contributing to it remain poorly understood \cite{Schneemann2006,Ranson2010}. This is mainly due to the pairing of bases along the backbone, which gives rise to the secondary or folded structures of RNA. Figure \ref{figure1} shows the cryo-EM images of three viruses: (a) Sindbis \cite{Zhang2002Sindbis}, (b) Flock house virus (FHV) \cite{Tsuruta1998FHV} and (c) Satellite tobacco mosaic virus (STMV) \cite{Schroeder2014STMV}. The capsid structures of these viruses are well characterized but the profile of their RNAs has remained the focus of many studies \cite{Zhang2002Sindbis, Tsuruta1998FHV, Schroeder2014STMV,Zeng2012,Linger2004SindbisRNA,Lindenbach2002FHVRNA}. As illustrated in Fig.~\ref{figure1}, for Sindbis, RNA fills out the entire capsid while for STMV, RNA covers basically the capsid wall. In case of FHV, RNA forms a thick layer inside the capsid. 

To date all theoretical studies deciphering the profile of RNA inside the capsid have considered RNA either as a linear or a branched polymer \cite{Siber2008,Zandi2009,Siber2010,Paul2013,Erdemci2014,Wagner2015,Hagan,Singaram2015,Gonca2016,Venky2016,Bruinsma-Grosberg, Erdemci2016}. While treating RNA as a branched polymer is a good first step, it does not account for the presence of loops and for their entropies, and it is known that base-pairing gives rise to more complex structures like pseudoknots. Pseudoknots have a strong effect on the compaction of the molecule by pairing together far apart bases along the chain and making the structure globular and as such RNAs are significantly more compact than self-avoiding branched polymers \cite{Hyeon2006,Fang2011,Ben-Shaul2015}. 

Modeling RNA as a branched polymer, Lee and Nguyen find that there are two different scenarios for RNA profiles inside the capsid \cite{Lee-Nguyen2008}. If the charge density of the capsid is high, the genome mostly sits close to the wall, otherwise, due to entropic effects for the low capsid charge density, the RNA concentration is higher at the center of the capsid \cite{Lee-Nguyen2008}.

In this Letter, we introduce a new model and show that RNA can assume different profiles inside the capsid as the ones presented in Fig.~\ref{figure1}, almost independent of the strength of capsid-genome interactions. The model considers RNA base-pairing and its saturation, {\it i.e.,} one base can pair with at most one base at a time. We show that the pairing can lead to a transition from an extended to a compact globular structure, despite the saturation. This transition is sharp (first order) as a function of the  genome length but it is smooth (second order) as a function of the number of base-pairs (BPs) per unit length or the strength of interaction that depends on RNA primary sequence dictating its secondary and tertiary structures. 

In the collapsed regime, RNA mostly covers the surface of the capsid (Figs.~\ref{figure1}(c),~\ref{kink-solution}, S2 and S3 solid lines) whereas in the swollen regime it is uniformly distributed inside the capsid with a density slightly higher at the attractive wall (Figs.~\ref{figure1}(a),~\ref{kink-solution}, S2 and S3 dotted lines). We also obtain a kink type profile (Figs.~\ref{kink-solution}, S2 and S3 dashed red lines), associated with the first order transition. In this case, there are three distinct regions for RNA density inside the capsid. In the vicinity of the capsid wall, the genome density is very high, immediately followed by an intermediate density regime before the start of the kink. Towards the capsid center, the density is almost zero. This profile is similar to the RNA concentration presented in Fig.~\ref{figure1}(b).

Unlike the previous work \cite{Lee-Nguyen2008}, we find that the profile of RNA strongly depends on the genome base-pairing and length. While the genome-capsid electrostatic interaction contributes to the RNA conformation inside viral shells \cite{ElSawy2011}, here we show that regardless of the strength of capsid-genome interaction, all three profiles illustrated in Fig.~\ref{figure1} can be obtained for both high and low surface charge densities as a result of base-pairing (Figs.~\ref{kink-solution}, S2 and S3). Our findings are consistent with the experimental studies of CPs of CCMV with poly(styrene sulfonate) (PSS)--a negatively charged linear polymer--in which despite the strong capsid charge density, PSS fills out more or less the entire shell  \cite{Hu2008}. 

The model also considers the presence of previously ignored pseudoknots (see below). Experimental studies on several viruses confirm the presence of pseudoknots in viruses \cite{Zeng2012}. The role of pseudoknots in the RNA packaging by CPs has not been thoroughly investigated, but pseudoknots clearly make RNA more compact.


To obtain the profile of RNA inside the capsid, we consider a model, where the interaction energy $\varepsilon_{ij}(r)$ is independent of the nature of the bases $i$ and $j$ and denoted $\varepsilon$. 
This amounts to using an effective attractive interaction, averaged over several bases within one RNA persistence length \cite{Montanari2000}. This effective pairing interaction $\varepsilon$ could also be defined as the average pairing energy of the RNA over the whole sequence. Note that it has been found that the sequence of viral RNAs contains a larger fraction of BPs than non-viral ones \cite{Yoffe2008,Gopal,Luca2015,Bundschuh2002}. The corresponding $\varepsilon$ is thus larger in viral RNAs than in non-viral ones.

The pairing partition function of the model reads
\begin{align}
\label{partition0}
\mathcal{Z}&=\int \left \{ \prod_{i=1}^N d {\bf r}_i \right \} e^{-\frac{3}{2 a^2} \sum_i ({\bf r}_{i+1}-{\bf r}_i)^2-\frac{\upsilon_0}{2}\sum_{i,j}\delta({\bf r}_i-{\bf r}_j)} Z_N
\end{align}
with $a$ the Kuhn length and $\upsilon_0$ the excluded volume parameter. $Z_N$ is the contribution of the saturated base-pairing to the partition function. Performing a virial expansion, we can write
\begin{equation}
\label{ZV}
Z_N=1+\sum_{i<j}V_{ij}+\sum_{i<j<k<l} \left( V_{ij}V_{kl}+V_{ik}V_{jl}+V_{il}V_{jk} \right)+\cdots
\end{equation}
with $V_{ij}=v\delta({\bf r}_i-{\bf r}_j)$ and $v=e^{\beta\varepsilon}-1$ the Mayer function associated with the energy $-\varepsilon$ of a bond $(i,j)$. Note also that the strength of the pairing depends on the temperature, the salt concentration and the pH of the solution. {As noted above, it also depends on the number of BPs}.
 Based on Wick's theorem, it can be shown \cite{Orland2002,Vernizzi2005} that 
\begin{align}
\label{ZN}
Z_N =\frac{1}{\mathcal{N}}\int \prod_{i=1}^N d\phi_i e^{-\frac{1}{2}\sum_{i,j}\phi_iV_{ij}^{-1}\phi_j}\prod_{i=1}^N(1+\phi_i)
\end{align}
and $<\phi_i\phi_j>=V_{ij}= v\delta({\bf r}_i-{\bf r}_j)$.  
The expression in Eq.~\ref{ZV} reveals that $Z_N$ is the sum of all possible sets of base pairings, with a weight $v$ associated to each pairing. 
As such, it comprises both planar and pseudoknotted structures.  Consider for example the case with two pairings with indices $i<j<k<l$. The pairing $V_{il}V_{jk}$ represents a helical fragment while  $V_{ik}V_{jl}$ represents a pseudoknot (see the supplemental material (SM)).

Using standard methods of polymer physics \cite{deGennes1979} and introducing the new fields $\chi({\mathbf r})$ and $\vartheta({\bf r})$ (SM), the partition function can be written as
\begin{equation}
\label{partition4}
\mathcal{Z}=\int \mathcal{D} \chi   \mathcal{D} \vartheta e^{ -\frac{v}{2}\int d{\bf r} \chi^2({\bf r}){-\frac{\upsilon_0}{2}{\int d{\bf r} \vartheta^2({\bf r})}}} \int d{\bf r} d{\bf r'} {\cal Q}({\bf r}, {\bf r'})
\end{equation}
where ${\cal Q}({\bf r}, {\bf r'})= \langle {\bf r} | e^{-N \hat H} | {\bf r'} \rangle$ is the single chain partition function in an external field, and we use the Dirac notation for the evolution operator of the Schr\"odinger equation. The corresponding Hamiltonian is given by $\hat H = -\frac{a^2}{6} \nabla^2+ i\upsilon_0 \vartheta(r ) - \log(1+v \chi(r))$. In the limit of long chains $N \to \infty$, for collapsed or confined chains, we employ the ground state dominance approximation {\cite{deGennes1979,Borukhov1998,Siber2008}}. After standard calculations (SM), we find the free energy of the system $\beta{\mathcal{F}}=-\log\mathcal{Z}$ as

\begin{align} \label{FE2}
\beta{\mathcal{F}} &= \int \mathrm{d}{{\bf{r}}} \bigg{\{} \frac{a^2}{6} | \nabla   \psi({\bf r}) |^2 +f(\psi)-\lambda(\psi^2({\bf r})-\frac{N}{V})
  \bigg{\}} 
\end{align}
with
\begin{align} \label{FE2p}
f(\psi)&=\frac{1}{2} \upsilon_0  \psi^4({\bf r}) + \frac{1}{8 v}({\sqrt{1+4v\psi^2({\bf r})}-1})^2  \nonumber \\ & - \psi^2({\bf r})\log(\frac{1}{2}(1+\sqrt{1+4v\psi^2({\bf r})})) 
\end{align}
with $\psi^2({\bf r})$ the genome monomer density at position $\rm r$ and $\lambda$ a Lagrange multiplier fixing the number of monomers inside the capsid,  $N=\int d{\bf r} \psi^2({\bf r})$ \cite{Hone1988}. The gradient term in Eq.~\ref{FE2} represents the elastic energy of the RNA polymer chain, and the term proportional to $\upsilon_0$ is the excluded volume contribution. The two other terms in Eq.~\ref{FE2p} represent the local pairing energy. The variation of Eq.~\ref{FE2} with respect to $\psi({\bf r})$ gives,
\begin{equation}
\label{difeq}
\frac{a^2}{6}\nabla^2 \psi=-\lambda \psi + \upsilon_0 \psi^3-\psi\log(\frac{1}{2}(1+\sqrt{1+4v\psi^2({\bf r})}))
\end{equation}
that can be solved numerically to obtain $\psi(r)$. This method also allows us to calculate the number of BPs, see SM.

In order to find the encapsidation free energy, we consider a polymer trapped inside an adsorbing sphere. The free energy (Eq.~\ref{FE2}) then becomes
\begin{equation}
\label{FE3}
\beta{\mathcal{F}_{int}}=\beta{\mathcal{F}}-\gamma\beta a^3\int dS \psi^2({\bf r})
\end{equation}
The last term represents the interaction energy of RNA with the capsid surface,
with $\gamma$ the {(adsorption)} energy per unit area. Minimizing the free energy, Eq.~\ref{FE3}, with respect to $\psi({\bf r})$ gives the same equation as Eq.~\ref{difeq}, but subject to the BC, $(\hat{n}\cdot \nabla \psi =\kappa \psi)|_{s}$, with $\kappa^{-1}=\frac{1}{6\beta a \gamma}$ being a length representing the interaction strength between the RNA and the wall. Fig. S1 illustrates the encapsidation free energy, Eq.~\ref{FE3}, as a function of the RNA length. For large pairing strength $\beta \epsilon$, the confinement free energy does not increase for longer RNAs as it does for weaker pairing indicating more stable system with higher pairing strength (SM).  

\begin{figure}[ht]
  \includegraphics[width=6.75cm]{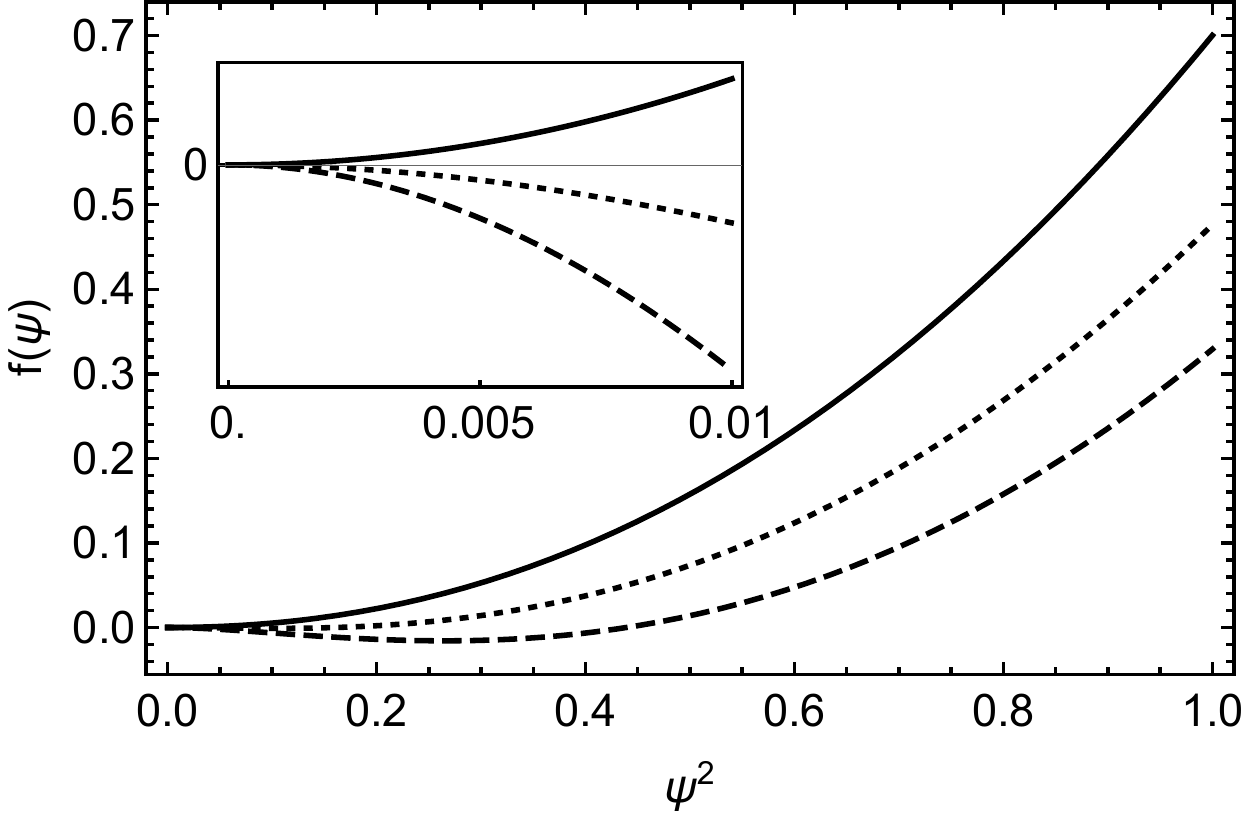}
 \caption{Free energy density as a function of RNA density for different base-pairing interaction energy values, $\beta\varepsilon=0.7$ (solid line), $\beta\varepsilon=1.3$ (dotted line) and $\beta\varepsilon=1.7$ (dashed line). Here $\upsilon_0=2$ $nm^3$ and inset shows the behavior around the origin. }
 \label{g-rho} 
\end{figure}

While the behavior of ${\mathcal{F}_{int}}$ vs $N$ is rather expected, a careful analysis of the ``bulk" free energy density (Eq.~\ref{FE2p}) indicates that there is a critical $\beta\varepsilon_c$ beyond which, the polymer will be in the collapsed phase. The plot of $f(\psi)$ vs $\psi^2$ in Fig.~\ref{g-rho} at a fixed RNA length shows a second order phase transition as a function of $\beta\varepsilon$. The effect is visible in the inset near the origin. For large $\psi^2$, contrary to the standard theta point for a polymer in a bad solvent, there is no need to include a repulsive 3-body interaction to avoid the collapse of the chain at infinite density. This is because of the saturation effect included in the model.

The profiles of RNA in the collapsed and extended states in the ``bulk" are illustrated in Fig.~\ref{f-r-N}.  To obtain the profiles we solve Eq.~\ref{difeq} in a spherical geometry. For the ``bulk" system we employ Dirichlet boundary conditions (BCs), $\psi(r=R)=0$, where $R$ is the radius of a sphere to be taken large enough to mimic the bulk. Dirichlet as well as Neumann conditions (both imply $\nabla \psi = 0$ on the boundary) guarantee that the monomer density is constant in the bulk.  Due to the nature of the symmetry we also have a BC at the center of the sphere as $\nabla \psi_{r=0}=0$. 

Figure~\ref{f-r-N} shows a plot of the monomer density vs the distance from the capsid center. As illustrated there is a phase transition: while a strong interaction ($\beta\varepsilon=5.0$, dot-dashed line) results in a collapsed phase, a weak interaction ($\beta\varepsilon=0.97$, dotted line) corresponds to a swollen phase, where the RNA spreads out uniformly. For larger radii, $R=40$ and $50$ $nm$, we obtained the same results indicating robustness of our findings. 

Using Eq.~\ref{difeq} with $(\hat{n}\cdot \nabla \psi =\kappa \psi)|_{s}$, we also obtained the RNA density inside an adsorbing sphere for both a strong ($ \beta\varepsilon=5.7$, dot-dashed line) and a weak interaction energies ($\beta\varepsilon=0.97$, dotted line). For a fixed genome length and capsid radius, in case of strong interaction, RNA mostly covers the surface of the capsid, whereas in the weak case, the genome essentially fills out uniformly the entire sphere. In the swollen regime the density is slightly higher at the wall, due to the attraction (Fig. \ref{f-r-N} inset, dotted profile), see below for the RNA profile as a function of its length and capsid-genome interaction.

\begin{figure}\label{f-r-N}
  \includegraphics[width=7cm]{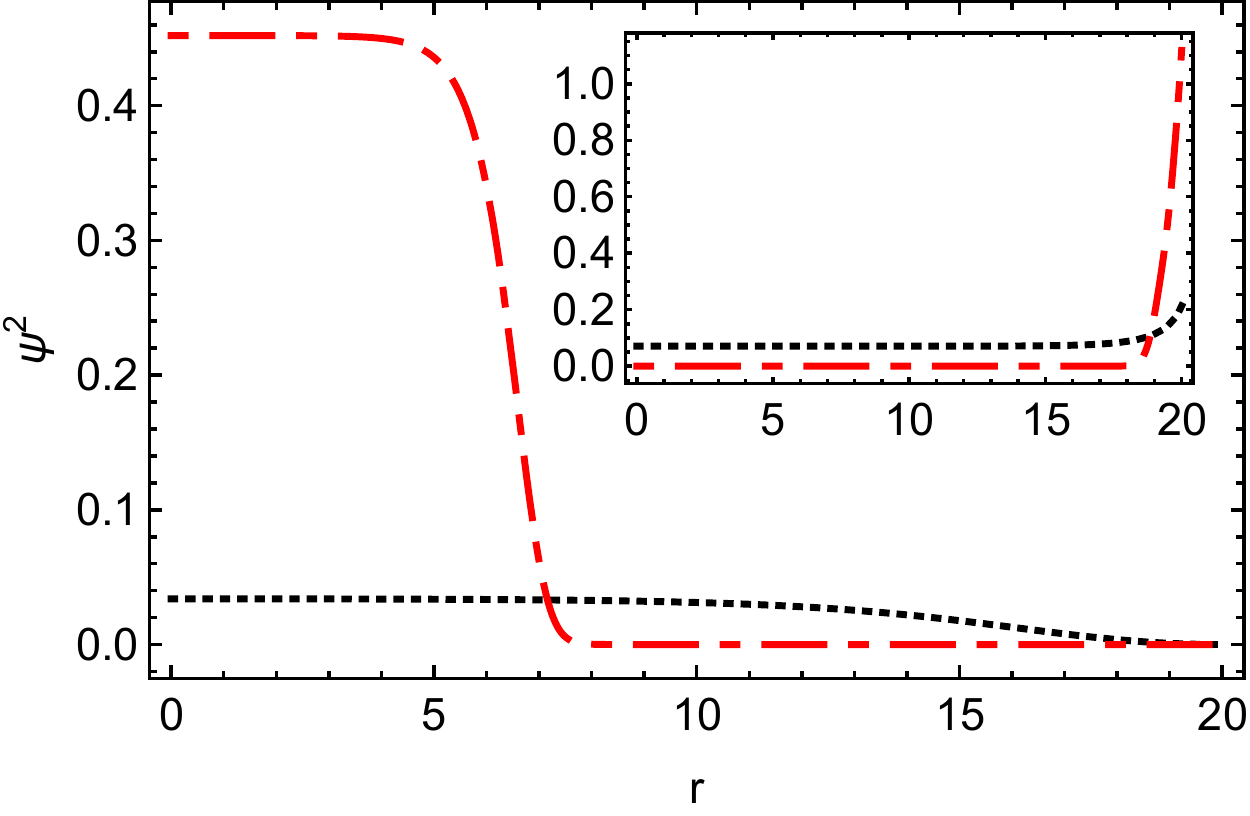}
 \caption{ RNA density profile for $N = 500$ with an interaction strength of $\beta\varepsilon =5.0$ (dot-dashed line) and $\beta\varepsilon=0.97$ (dotted line). The inset shows RNA density profile inside an adsorbing capsid for $N = 3000$ with $\beta\varepsilon=5.7$ (dot-dashed line) and $\beta\varepsilon=0.97$ (dotted line). Other parameters are $a=1$ $nm$, $\upsilon_0=2$ $nm^3$, $R=20$ $nm$ and $\kappa=0.5$ $nm^{-1}$.}
 \label{f-r-N} 
\end{figure}




We also studied the RNA profile inside viral shells as a function of its length for a fixed capsid-genome interaction ($\kappa$) and base-pairing strength ($\beta\varepsilon$). Figure \ref{kink-solution} illustrates the impact of the RNA length on its density profile, for a fixed value of $\beta\varepsilon=1.7$. A relatively short chain $N=525$ (solid line) covers entirely the inner surface, leaving the interior of the shell empty (collapsed phase, Fig. \ref{kink-solution} solid line). The longer chains are swollen throughout the shell with an increased density next to the wall (swollen phase, Fig. \ref{kink-solution} dotted and dot-dashed lines). Quite interestingly, we observe a domain wall or kink (a sharp drop from one value of minimum free energy to another minimum value of free energy \cite{Chaikin, Sanati1999}) between the swollen and collapsed phases for $N=2200$, Fig. \ref{kink-solution} dashed line. A simple scaling argument \cite{Chaikin} shows that the width of the kink region is {$\ell^*\sim\sqrt{\frac{a^2}{6}\frac{(\psi_{1}-\psi_{2})^2}{E_b}}$ where $\psi_{1}$ and $\psi_{2}$ are the density fields of RNA at the minima of free energies (see below) and $E_b$ the height of the energy barrier between the two minima}. The width $\ell^*$ matches very well our numerical solutions. Note we find all RNA profiles shown in Fig.~\ref{kink-solution} for both strong and weak capsid-genome interactions (SM).

\begin{figure}
  \includegraphics[width=7cm]{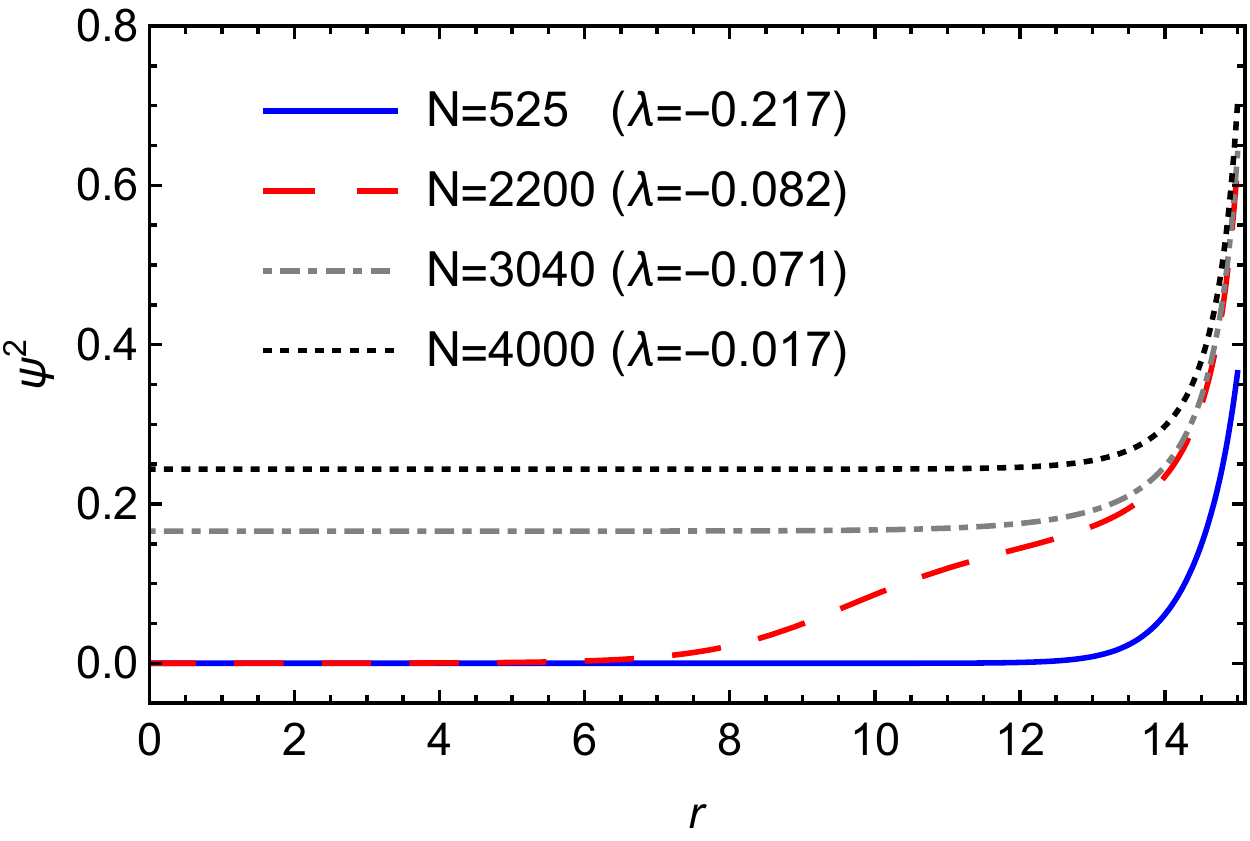}
 \caption{RNA density profile as a function of distance from the center of the shell for different chain lengths $N=525$ (solid line), $N=2200$ (dashed line), $N=3040$ (dot-dashed line) and $N=4000$ (dotted line). Other parameters are $\beta\varepsilon=1.7$, $a=1$ $nm$, $\upsilon_0=2$ $nm^3$, $R=15$ $nm$ and $\kappa=1.0$ $nm^{-1}$.}
 \label{kink-solution} 
\end{figure}

\begin{figure}
  \includegraphics[width=7cm]{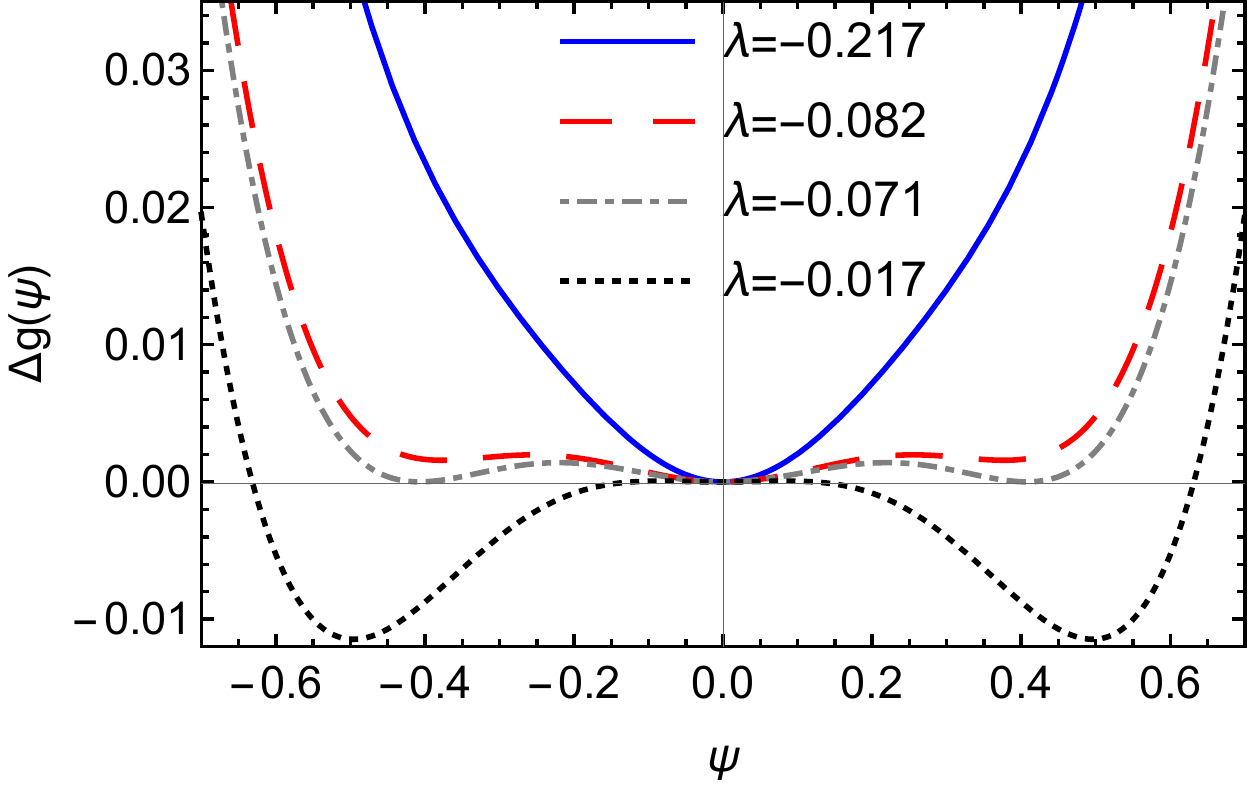}
 \caption{Energy density as a function of RNA density field $\psi$ for different chain lengths at $\lambda=-0.217$ (solid curve), $\lambda=-0.082$ (dashed curve), $\lambda=-0.071$ (dot-dashed curve) and $\lambda=-0.017$ (dotted curve). Other parameters are $\beta\varepsilon=1.7$, $a=1$ $nm$, and $\upsilon_0=2$ $nm^3$. The $\lambda$ values here correspond to the $N$ values presented in Fig.~\ref{kink-solution} (see the text).}
 \label{energy-density} 
\end{figure} 

To better understand the kink behavior, we examine the global and local minima of the Gibbs free energy density (Eq.~\ref{FE2p}) in the bulk, $\Delta g(\psi)=f(\psi) -\lambda \psi^2$.
Figure \ref{energy-density} is a plot of $\Delta g(\psi)$ vs $\psi$, corresponding to the $N$ (or $\lambda$) values given in Fig.~\ref{kink-solution}. As illustrated in Fig.~\ref{energy-density}, at $\lambda=-0.082$, there are two minima at $\psi=0$ and $\psi=~0.4$, which correspond to the RNA densities ($\psi^2=0$ and $\psi^2=0.16$) in different sides of kink ($\approx r=6$ and $r=13$) in Fig.~\ref{kink-solution}, the dashed line. For $N=4000$ (Fig.~\ref{kink-solution}), the interior density is fixed and is equal to the square of the minimum of the corresponding free energy ($\psi=0.5$) (Fig. \ref{energy-density}, dotted curve). Similarly, the solid curve in Fig. \ref{energy-density} has a minimum at $\psi=0$ and the corresponding density profile (Fig.~\ref{kink-solution}, solid line) is such that the density is constant and zero in the interior of the sphere. The density increases at the wall due to the attractive interaction between the RNA and the wall for all profiles. For $N=3040$ (Fig. \ref{kink-solution}), the interior density is fixed to one of the three degenerate minima of the free energy (Fig. \ref{energy-density}, dot-dashed curve) in agreement with the first order phase transition in the system. 

{We emphasize again that despite an intense ongoing research,} the precise profile of RNA in many viruses including one of the most studied viruses, CCMV is not yet clear. A careful analysis of many Cryo-EM images suggests that the density of RNA in the interior of CCMV capsid is almost zero and RNA is sitting in the vicinity of the capsid wall \cite{Fox1998CCMV}, similar to our results in Fig.~\ref{kink-solution} (dashed line). Quite interestingly, Cryo-EM images of virus-like-particles (VLPs) built from CCMV CPs and PSS molecules with similar length as the native CCMV RNA show that PSS almost completely fills out the capsid {\cite{Hu2008}. Our results are consistent with these experiments indicating that for a given capsid charge density and chain length, it is the base-pairing that defines the profile and stability of VLPs.}

In summary, we introduced a simple model to explore the profile of RNA inside viral shells. We showed that there is a critical base-pairing strength $\beta \epsilon_c$, below which RNA is in an extended state and almost uniformly fills out inside the capsid. For the strength larger than $\beta \epsilon_c$, RNA will be in a collapsed state and sits tightly next to the wall inside the capsid.  We note again that the strength of interaction here is defined as the average number of BPs per unit length. If an RNA is designed to have more BPs per unit length, it will have a larger $\beta\epsilon$. Furthermore, we found that for a given capsid-genome attractive interaction ($\gamma$) and the base-pairing strength $\beta \epsilon$, the profile of RNA could go through a sharp transition (first order) from collapsed to an extended state as a function of length of RNA. We observed a kink-type profile (remnant of the first order transition) revealing that there are three distinct density regions inside the capsid (Fig.~\ref{kink-solution}), a profile never captured in previous models. While the size and charge density of the capsids in Fig.~\ref{figure1} are different and the electrostatic interactions are important, our goal in this Letter was to single out the impact of base-pairing on the profile of RNA. We showed that all three profiles are possible regardless of the capsid size and charge density. Our work shows that a slight change in the number of RNA base-pairs and/or length can result in drastic changes in the profile of RNA from a collapsed to an extended form. Designing the appropriate RNA primary sequence, which defines the number of base-pairs and thus the behavior of RNA inside the shell, we can control the encapsidation efficiency and stability of VLPs for various material science and biological applications.
\\~\\ 
The authors would like to thank Jef Wagner and Siyu Li for useful discussions. This work was supported by the National Science Foundation through Grant No. DMR-1310687 (R.Z.).
\\~\\ 

\bibliography{rna}

\pagebreak
\onecolumngrid
\begin{center}
\textbf{\large RNA Base Pairing Determines the Conformations of RNA Inside Spherical Viruses
Supplemental Material}
\end{center}
\twocolumngrid
\setcounter{equation}{0}
\setcounter{figure}{0}
\setcounter{table}{0}
\setcounter{page}{1}

\renewcommand{\theequation}{S\arabic{equation}}
\renewcommand{\thefigure}{S\arabic{figure}}
\renewcommand{\thetable}{S\arabic{table}}
\renewcommand{\bibnumfmt}[1]{[S#1]}
\renewcommand{\citenumfont}[1]{S#1}

\section{Supporting Figures}

\begin{figure}[h]
  \includegraphics[width=7cm]{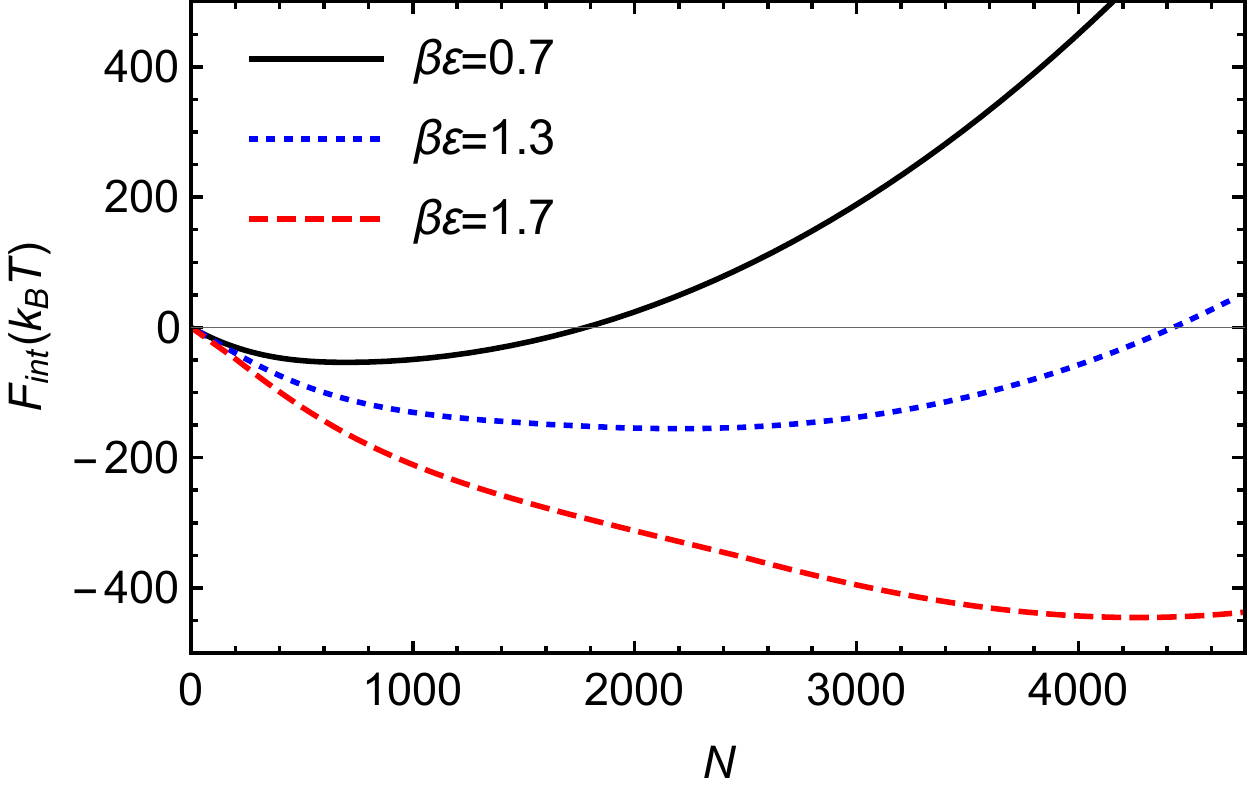}
 \caption{Encapsidation energy as a function of RNA length for three different base-pairing strengths $\beta\varepsilon=1.7$ (dashed line), $\beta\varepsilon=1.3$ (dotted line) and $\beta\varepsilon=0.7$ (solid line). Other parameters are $a=1$ $nm$, $\upsilon_0=2$ $nm^3$, $R=15$ $nm$ and $\kappa=1.0$ $nm^{-1}$.}
 \label{FvsN} 
\end{figure}

Figure~\ref{FvsN} illustrates the encapsidation free energy, Eq. 8, as a function of the RNA length for different base-pairing strengths $\beta\varepsilon=1.7$ (dashed line), $\beta\varepsilon=1.3$ (dotted line) and $\beta\varepsilon=0.7$ (solid line). As illustrated in the figure, for large pairing strength $\beta\epsilon$, the confinement free energy does not increase for longer RNAs as it does for weaker pairing. In other words, as the pairing strength increases, the minimum of the free energy moves towards longer chains and becomes deeper indicting a more stable system.

\begin{figure}[h]
  \includegraphics[width=7cm]{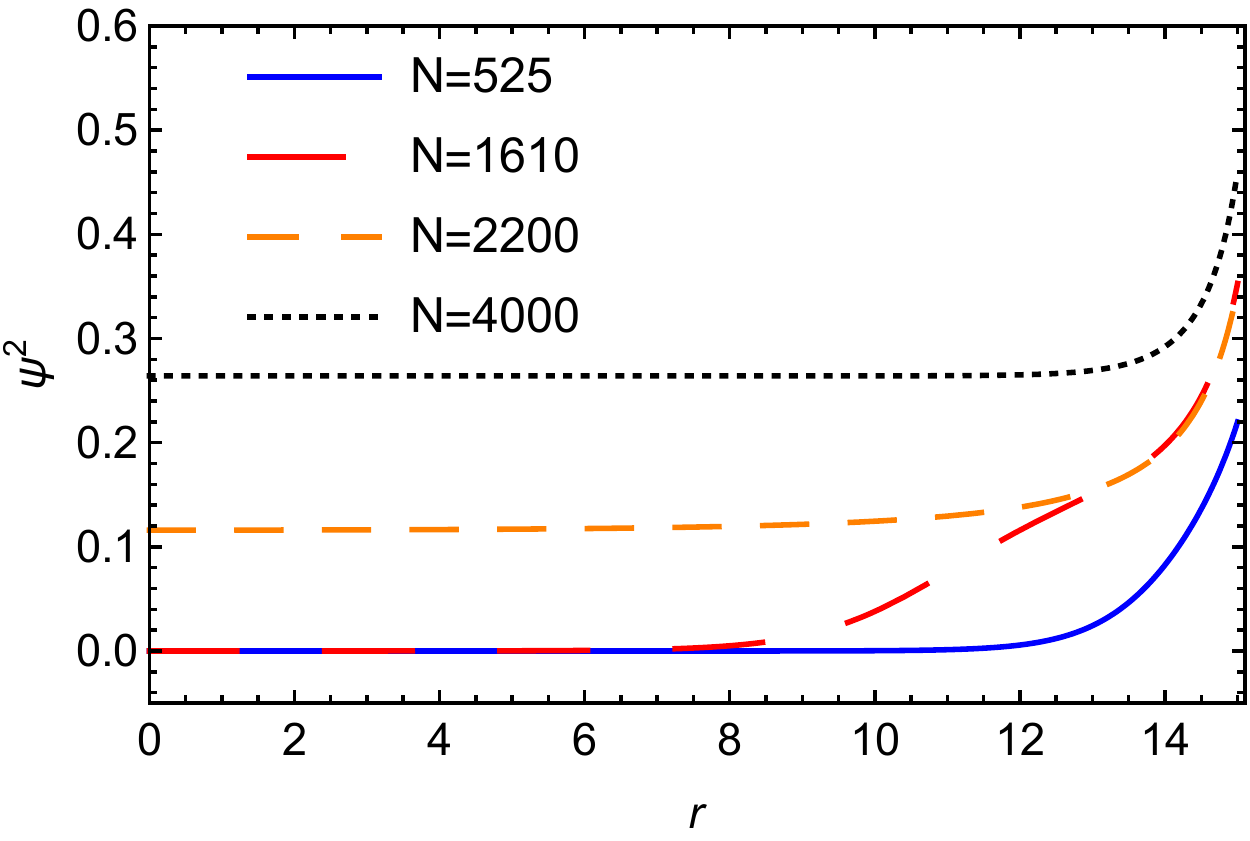}
 \caption{\footnotesize RNA density field as a function of distance from the center of the shell for different chain lengths $N=525$ (solid blue line), $N=1610$ (dashed red line), $N=2200$ (short-dashed orange line) and $N=4000$ (dotted black line). Other parameters are $\beta\varepsilon=1.7$, $a=1$ $nm$, $\upsilon_0=2$ $nm^3$, $R=15$ $nm$ and $\kappa=0.5$ $nm^{-1}$.}
 \label{kink-solution05} 
\end{figure}

\begin{figure}[h]
  \includegraphics[width=7cm]{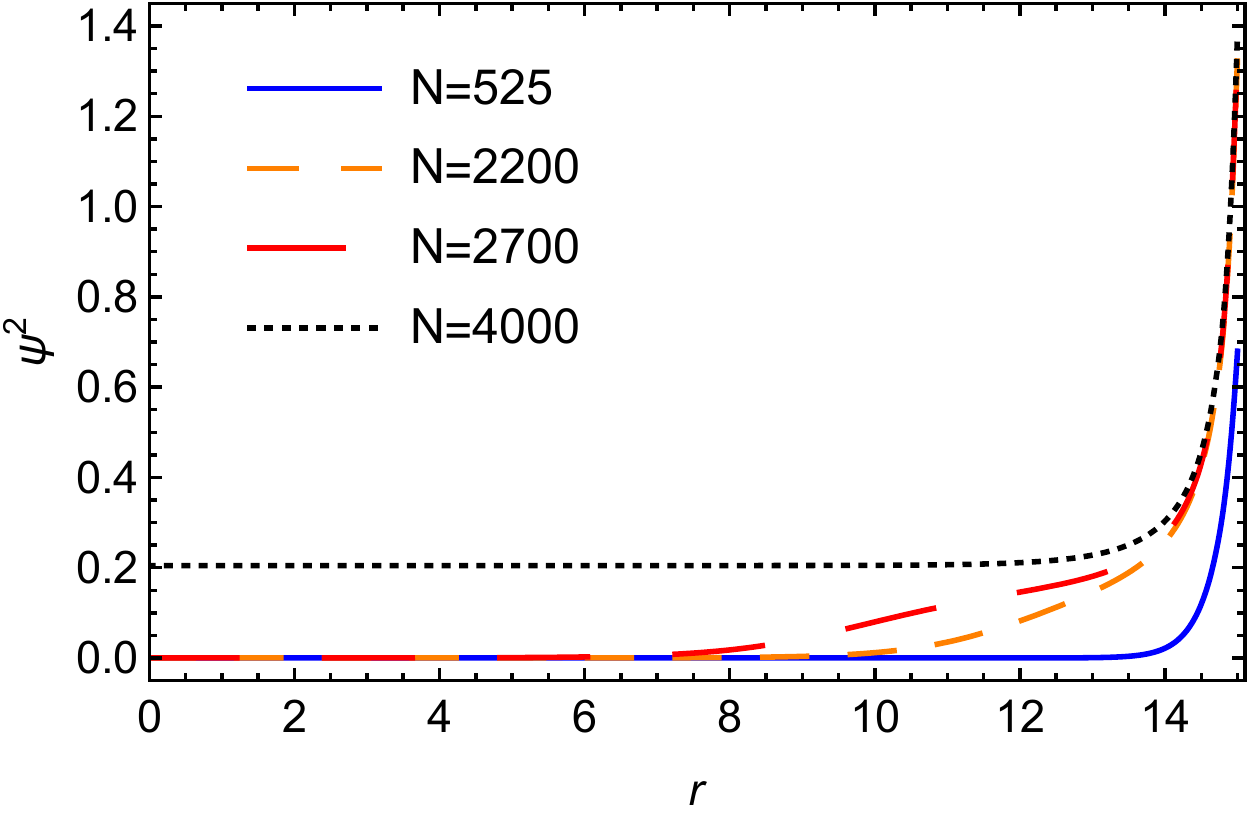}
 \caption{\footnotesize RNA density field as a function of distance from the center of the shell for different chain lengths $N=525$ (solid blue line), $N=2200$ (short-dashed orange line), $N=2700$ (dashed red line) and $N=4000$ (dotted black line). Other parameters are $\beta\varepsilon=1.7$, $a=1$ $nm$, $\upsilon_0=2$ $nm^3$, $R=15$ $nm$ and $\kappa=2.0$ $nm^{-1}$.}
 \label{kink-solution2} 
\end{figure}

Figures 4, S2 and S3 illustrate the impact of the strength of the genome-capsid attractive interaction on RNA profiles for $\kappa=1$, $\kappa=0.5$ and $\kappa=2.0$, respectively.  Each of these figures shows the behavior of RNA inside a viral shell as a function of its length for a fixed capsid-genome interaction ($\kappa$) and base-pairing strength $\beta\varepsilon$. Figures~\ref{kink-solution05} and \ref{kink-solution2} show the impact of the length of RNA on its density profile inside the capsid for a stronger $\kappa=2.0$ and a weaker capsid-genome interactions $\kappa=0.5$, respectively.  A relatively short chain (solid blue lines) covers entirely the inner surface, leaving the interior of the shell empty. The longer chain is extended throughout the shell with an increased density next to the wall (dotted black lines). There is also a domain wall (kink) between the swollen and collapsed phases (dashed red lines). We emphasize that all three RNA profiles (collapsed, swollen and kink) are observed because of changes in the length of the encapsidated RNA regardless of the capsid-genome interaction, as revealed in Figs~\ref{kink-solution05}, \ref{kink-solution2} and Fig.~4 in the text.

\section{Derivation of Equations 4 and 5 in the text}

Equation 3 in the main text can be rewritten as, 

\begin{align}
\label{ZN}
Z_N &=\frac{1}{\mathcal{N}}\int \prod_{i=1}^N d\phi_i e^{-\frac{1}{2}\sum_{i,j}\phi_iV_{ij}^{-1}\phi_j}\prod_{i=1}^N(1+\phi_i)\nonumber\\
&=\int \prod_{i=1}^N d\phi_i \prod_{i=1}^N d\xi_i e^{-\frac{1}{2}\sum_{i,j} \xi_i V_{ij}{\xi_j}+i\sum_i\phi_i\xi_i}\prod_{i=1}^N(1+\phi_i)
\end{align}
where $\xi_i$ is introduced in the Gaussian integral.  The exponential of the interaction terms in Eq.~\ref{ZN} can be written as
\begin{align}
\label{exp_int}
e^{-\frac{1}{2}\sum_{i,j} \xi_i V_{ij}{\xi_j}}&=e^{-\frac{1}{2} v \int d{\bf r} (\sum_i \xi_i \delta({\bf r}-{\bf r}_i))^2}\nonumber\\
&=\int \mathcal{D} \chi e^{-\frac{v}{2}\int d{\bf r} \chi^2({\bf r})-iv \sum_i \xi_i \chi({\bf r}_i)}
\end{align}
with the introduction of the auxiliary field $\chi({\mathbf r})$ in terms of the Gaussian integration. Plugging Eq.~\ref{exp_int} into Eq.~\ref{ZN} gives
\begin{align}
\label{ZN2}
Z_N = &\int \prod_{i=1}^N d\phi_i \prod_{i=1}^N d\xi_i  \nonumber\\
&\int \mathcal{D} \chi e^{-\frac{v}{2}\int d{\bf r} \chi^2({\bf r})+i\sum_i(\phi_i-v \chi({\bf r}_i))\xi_i} \prod_{i=1}^N(1+\phi_i) \nonumber\\
&= \int \mathcal{D} \chi e^{-\frac{v}{2}\int d{\bf r} \chi^2({\bf r})} \prod_{i=1}^N(1+v \chi({\bf r}_i))
\end{align}
This new form of $Z_N$ can be inserted in the partition function given in Eq.~1 in the paper and then we have,
\begin{align}
\label{Z2}
\mathcal{Z}= & \int \prod_{i=1}^N d {\bf r}_i e^{-\frac{3}{2 a^2} \sum_i ({\bf r}_{i+1}-{\bf r}_i)^2-\frac{\upsilon_0}{2}\sum_{i,j}\delta({\bf r}_i-{\bf r}_j)}  \nonumber\\
& \int \mathcal{D} \chi e^{-\frac{v}{2}\int d{\bf r} \chi^2({\bf r})} \prod_{i=1}^N(1+v \chi({\bf r}_i))
\end{align}
Here the two-body excluded volume interaction can be evaluated as
\begin{align}
\label{part_exc}
e^{-\frac{\upsilon_0}{2}\sum_{i,j}\delta({\bf r}_i-{\bf r}_j)}& =e^{-\frac{\upsilon_{0}}{2}\int d{\bf r} \hat{\rho}_m^2}\nonumber\\
 &=\int \mathcal{D} \vartheta e^{-\frac{\upsilon_0}{2}{\int d{\bf r} \vartheta^2({\bf r})}-i \upsilon_0\int d{\bf r}\hat{\rho}_m\vartheta({\bf r})}\nonumber\\
&=\int \mathcal{D} \vartheta e^{-\frac{\upsilon_0}{2}{\int d{\bf r} \vartheta^2({\bf r})}-i\upsilon_0\sum_i\vartheta({\bf r}_i)}
\end{align}
where we have introduced the field $\vartheta({\bf r}) $ conjugate to the polymer density field $\hat{\rho}_m=\sum_{i} \delta({\bf r}-{\bf r_i})$. Using Eq.~\ref{part_exc},
in the continuous limit, the partition function becomes
\begin{eqnarray}
\label{partition3}
\mathcal{Z}&=&\int \mathcal{D} \chi   \mathcal{D} \vartheta e^{ -\frac{v}{2}\int d{\bf r} \chi^2({\bf r}){-\frac{\upsilon_0}{2}{\int d{\bf r} \vartheta^2({\bf r})}}} \nonumber \\
&\times& \int {\cal D} {\mathrm r}(s) e^{- H(r(s))} \nonumber \\
\end{eqnarray}
with
\begin{align}
H(r(s))=& \int_0^N ds \bigg{(} \frac{3}{2 a^2} \big{(}\frac{d{\mathrm r}(s)} {ds} \big{)}^2 \nonumber\\
&+ i\upsilon_0 \vartheta(r(s))- \log(1+v \chi(r(s))) \bigg{)}
\end{align}
The chain part of the partition function can also be rewritten as
\begin{equation}
\int {\cal D} {\mathrm r}(s) e^{- H(r(s))} = \int d{\bf r} d{\bf r'} {\cal Q}({\bf r}, {\bf r'})
\end{equation}
which together with Eq.~\ref{partition3} yields Eq.~4. From its definition,  ${\cal Q}({\bf r}, {\bf r'})= \langle {\bf r} |e^{-H}| {\bf r'} \rangle$ satisfies the Schr\"odinger  equation and assuming ground-state dominance, we have
\begin{eqnarray}
\label{ground}
&&\int d{\bf r} d{\bf r'} {\cal Q}({\bf r}, {\bf r'})
\approx_{N \to \infty} \nonumber \\
&{\rm Min}_{\{\psi({\bf r})\}}& \exp {\Bigg(-\int d{\bf r} \Big( \frac{a^2}{6} (\nabla  \psi({\bf r}))^2 +i\upsilon_0 \vartheta({\bf r})  \psi^2({\bf r})}, \nonumber \\
&&{-\log(1+v \chi({\bf r})) \psi^2({\bf r})  - \lambda (\psi^2({\bf r}) - \frac{N}{V}) \Big) \Bigg)} \nonumber \\
\end{eqnarray}
where $\psi({\bf r})$ is the eigenfunction representing the monomer density field. Inserting Eq. \ref{ground} into Eq. 4 and integrating out the $\vartheta$ field, the partition function simply becomes
\begin{equation}
\mathcal{Z}= \int \mathcal{D} \chi \exp^{-\beta{\mathcal{F}}}
\end{equation}
with the free energy 
\begin{multline} \label{FE}
\beta{\mathcal{F}} = \int \mathrm{d}{{\bf{r}}} \bigg{\{} \frac{a^2}{6} | \nabla   \psi({\bf r}) |^2
 +\frac{1}{2} \upsilon_0  \psi^4({\bf r})  + \frac{1}{2} v\chi^2({\bf r}) \\ - \psi^2({\bf r})\log(1+v \chi({\bf r})) -\lambda(\psi^2({\bf r}) -\frac{N}{V}) \bigg{\}}
\end{multline}
where $\psi^2({\bf r})$ is the genome monomer density at position $\rm {\bf r}$ and $\lambda$ is a Lagrange multiplier fixing the number of monomers inside the capsid,  $N=\int d{\bf r} \psi^2({\bf r})$. 

The variation of Eq.~\ref{FE} with respect to the field $\chi$ results in $\chi=\frac{\sqrt{1+4v\psi^2}-1}{2 v}$, which when plugged back into Eq.~\ref{FE}, yields Eq. 5 of the main text,
\begin{align} \label{FE2}
\beta{\mathcal{F}} &= \int \mathrm{d}{{\bf{r}}} \bigg{\{} \frac{a^2}{6} | \nabla   \psi({\bf r}) |^2 +f(\psi)-\lambda(\psi^2({\bf r})-\frac{N}{V})
  \bigg{\}}.
\end{align}

\section{The number of base-pairs}
The model presented in the text allows us to calculate the average number of base-pairs as follows,
\begin{align}
\label{Npb}
N_{pb}&=\frac{\partial \log (\mathcal{Z})}{\partial \beta\epsilon}=-e^{\beta \epsilon}\frac{\partial{(\beta \mathcal{F})}}{{\partial v}}\\ \nonumber
&=\frac{e^{\beta\epsilon}}{4 v^2} \int \mathrm{d}{{\bf{r}}}\; [1+2v\psi^2({\bf r})-\sqrt{1+4v\psi^2({\bf r})})].
\end{align}

For the profile presented in Fig. 3, based on Eq.~\ref{Npb}, the number of base-pairs are $N_{pb}=9$ and $N_{pb}=217$ for $\beta\varepsilon=0.97$ and $\beta\varepsilon=5.0$, respectively. As the strength of base-pairing $(\beta\varepsilon)$ increases, the number of base-pairs increases too. For the inset of Fig. 3, the length of RNA is N = 3000 and the number of base-pairs are $N_{pb} = 1350$ and $N_{pb} = 180$ for the case of strong and weak pairing interactions. We also calculated the number of base-pairs for each profile in Fig. 4. As the number of encapsidated monomers increases, the ratio of number of base-paired to the total monomer number increases. For $N=525, 2200, 3040$ and $4000$, the ratio of $N_{pb}/N=0.162,0.189,0.194,0.213$, respectively.

\section{Pseudoknots}
\begin{figure}[h]
  \includegraphics[width=8.7cm]{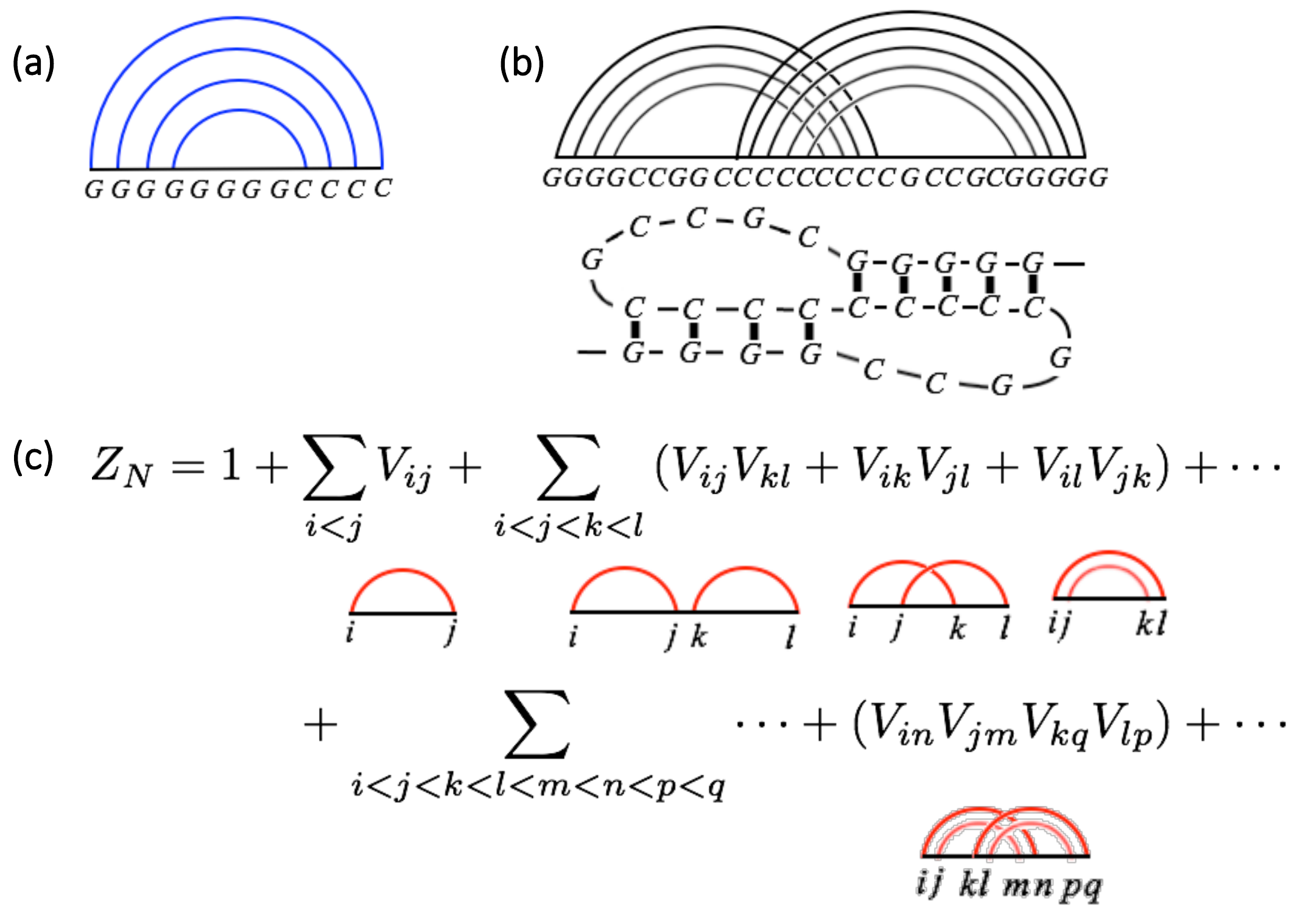}
 \caption{(a) Arc representation of the secondary structure of RNA. (b) Diagram with crossing arcs and structure of an RNA pseudo-knot. Crossing arcs indicate the presence of a pseudo-knot. (c) Equation 2 of the main text and the appearance of pseudo-knots in the sum.}
 \label{pseudoknots2} 
\end{figure}

The model allows pairing of far apart bases along the chain giving rise to pseudoknots. While the appearance of pseudoknots might not be obvious {in the expression} for free energy, it is more transparent in the partition function (Eq. 2), which considers RNA base-pairing with ``saturation", i.e., one base can interact with at most one base at a time. Figure \ref{pseudoknots2} illustrates the arc diagram representation of the interaction terms in Eq. 2 and how the formation of pseudoknots is included in the model.

\end{document}